\documentclass{article}

\usepackage{arxiv}

\usepackage[utf8]{inputenc} % allow utf-8 input
\usepackage[T1]{fontenc}    % use 8-bit T1 fonts
\usepackage{hyperref}       % hyperlinks
\usepackage{url}            % simple URL typesetting
\usepackage{booktabs}       % professional-quality tables
\usepackage{amsfonts}       % blackboard math symbols
\usepackage{nicefrac}       % compact symbols for 1/2, etc.
\usepackage{microtype}      % microtypography
\usepackage{lipsum}
\usepackage{graphicx}
\usepackage{subfigure}
\usepackage{xcolor}
\usepackage[ruled,linesnumbered,vlined]{algorithm2e}

\title{High Performance Consensus without Duplication: Multi-pipeline Hotstuff}

\author{
 Taining Cheng \\
  School of Software\\
  Yunnan University\\
  Kunming, China\\
  \texttt{tirning@outlook.com} \\
}

\begin{document}
\maketitle
\begin{abstract}
The state-of-the-art HotStuff operates an efficient pipeline in which a stable leader drives decisions with linear communication and two round-trips of message. However, the unifying proposing-voting pattern is not sufficient to improve the bandwidth and concurrency performance of the modern system. In addition, the delay corresponding to two rounds of message to produce a certified proposal in that scheme is a significant performance bottleneck. Thus, this study developed a new consensus protocol, Multi-pipeline HotStuff, for permissioned blockchain. To the best of the authors’ knowledge, this is the first protocol that combines multiple HotStuff instances to propose batches in order without a concurrent proposal, such that proposals are made optimistically when a correct replica realizes that the current proposal is valid and will be certified by quorum votes in the near future. Because simultaneous proposing and voting are allowed by the proposed protocol without transaction duplication, it produced more proposals in every two rounds of messages. In addition, it further boosted the throughput at a comparable latency with that of HotStuff. The evaluation experiment conducted confirmed that the throughput of Multi-pipeline HotStuff outperformed that of the state-of-the-art protocols by approximately 60\% without significantly increasing end-to-end latency under varying system sizes. Moreover, the proposed optimization also performed better when it suffers a bad network condition.
\end{abstract}

% keywords can be removed

\section{Introduction}
The emergence of blockchain has significantly affected Byzantine fault-tolerant (BFT) consensus in the last decade. As a core infrastructure of blockchain, the BFT consensus protocol \cite{10.1145/98163.98167,10.1145/3335772.3335936} build the basis for implementation of state machine replica and the execution of smart contracts in the decentralized network. Through the BFT protocol, each party of the system maintains the same state of data, even enduring arbitrary failures when a request that can change the internal state of the system originates from outside. The protocol ensures that $2f+1$ out of $n$ correct parties agree on the order of requests (where $n$ is system size and $f$ is the threshold of failures, such that $n>3f$). Subsequently, the state transition of the system occurs through the execution of requests in the same order. Recently, this idea was used to build a blockchain service, a modern cryptocurrency system, in addition to the replication system. To reach a consensus on the order of requests, the classical BFT protocol employs a leader-based method to drive consensus instances to continuously maintain the state of the system under partial synchrony \cite{10.1145/42282.42283}, where the message delay is at most $\Delta$ after an unknown global stabilization time (GST). Futher, in PBFT  \cite{10.5555/296806.296824}, each consensus decision amortizes the authenticator cost of the network, $O(n^2)$, with the cost of view-change being more, namely $O(n^3)$.

However, significant costs hindered the large-scale deployment of the system until the HotStuff is presented, which is a new paradigm that allows optimistic responsiveness \cite{10.1007/978-3-319-78375-8_1} and linear view-change. With threshold signature schema \cite{10.1007/3-540-45539-6_15}, the quadratic cost of the normal case can be reduced to a linear form, and an additional consensus phase overcomes a stumbling block in BFT consensus. Consequently, each block acquires the first two phases to lock a safe position that will never be forked in the future and an additional phase to commit the block in that position. However, the simple addition of a phase is insufficient. Thus, a chained structure wherein pipeline voting results in the following proposal is introduced. This structure unifies three phases of one block into a proposing-voting pattern, effectively improving system performance with the features of a linear cost and responsiveness. Certain organizations, such as Meta, have adopted it to implement permissioned blockchain services.

Although the introduction of the chained structure opened a new chapter that bridged the classical BFT and blockchain consensus. a bottleneck gradually surfaced. In particular, in the context of this linear proposing-voting pattern, it is referred to as a single-pipeline bottleneck in this paper, which describes the poor efficiency when it is used as a critical commit path in the consensus protocol. In the proposing-voting pattern, a replica can have more than one role. The leader collects quorum votes to generate QC when the validator sends its vote of the proposal to it. Thereafter, a proposal extends the block associated with that QC. From a macro perspective, regardless of the role, either a proposal drives replicas to validate the proposal and send a vote, or quorum votes drives a new leader to propose. Investigations of the manner in which driven messages are produced and transmitted can entail more space to break through the performance bottleneck.

\textbf{The performance bottleneck}. First, a simple proposing-voting pipeline results in the replica always facing a blank period of time that requires no action, owing to no driven message being received. With the system operation, the accumulation of unused blank time limits the performance of a system that cannot facilitate execution. Second, the pipelined operation provides the ability that extends an on-consensus block before being committed. This is in contrast to only extending a decided block based on safety property such as classical BFT. Thus, the system can now generate a new block after a proposing and voting cycle and consequently commit a block. However, But the delay of two rounds of messages remains essential. Third, every party requires considerable time to perform cryptographic primitives and hash function, such as producing signatures and mapping arbitrary-size input to a fixed-size output, each of which is computationally intensive. Meanwhile, the transmission module is free when the replica performs computation on the CPU, thereby resulting in the concurrent capability of the multi-core system and the bandwidth not being fully exerted. Consequently, the single proposing-voting approach can be significantly optimized.

Therefore, the natural intuition for optimizing the single pipeline pattern involves each replica producing the maximum blocks possible to fully drain the bandwidth and the ability of concurrent \cite{stathakopoulou2019mir,10.1145/2976749.2978399}. Some solution was built around this idea was proposed, allowing parallel leaders to propose batches independently and concurrently, such as incase of Mir BFT \cite{10.1145/2976749.2978399}. This led to improved performance, despite the need for an extra transaction partition strategy to preclude duplication problems and duplication attacks, which are caused through independent proposing. While the dilemma is that these duplications can simply be filtered out after an ordering, but the damage has already been done---excessive resources, bandwidth, and possibly CPU have been consumed, although the duplication is a consequence of insufficient optimization. Consequently, a natural question arises: Is it possible to design a consensus protocol that avoids duplication while retaining the ability to overcome the performance bottleneck made by the single pipeline pattern?

\textbf{Contributions.} To the best of the authors' knowledge, Multi-pipeline HotStuff (MPH), which is proposed in this paper, is the first novel BFT protocol that ensures that one proposal is generated per round of messages. Without increasing the latency, MPH produces twice as many proposals as those produced by HotStuff owing to the redesigning of the consensus instance, which combines two HotStuff instances to drive consensus. Moreover, MPH provides efficient view-change and responsiveness \cite{10.1007/978-3-319-78375-8_1} through the adoption of a three-phase proposing-voting pattern \cite{10.1007/978-3-662-46803-6_10}. Thus, each block is committed by its third QC and locked by its second QC. 

The conservative approach of HotStuff involves the extension of the block that has a QC associated with it. In contrast, the key idea involved in MPH is that the correct replica optimistically extends a block based on a verified (not certified) proposal when it realizes that the proposal can be voted on, and it becomes the leader of the next view. Consequently, this study redesigned a new view mechanism in MPH, which only comprised one round of messages, with the replicas always waiting for votes and blocks. This optimistic extension resolved several of the challenges mentioned above. First, simultaneous proposing and voting can facilitate concurrency when a replica can perform both roles. The multi-pipeline pattern allows a new proposal to be made in each round of messages and for one proposal to be voted on in each round of messages. In addition, the blank period of time in the single pipeline owing to no driven messages being received in the single pipeline is eliminated. Second, the optimistic proposal by the correct leader reduces the two rounds of delay to one. When a replica is voting and preparing a proposal, the computationally intensive operations and network transmission are fully exploited. Third, MPH does not suffer from the duplication issue because it does not adopt parallel leaders to address the performance bottleneck.

The challenge in MPH is ensuring the total order of all proposals that are optimistically proposed. Consider a case wherein replicas have received no messages and timeouts, and two blocks are not certified because neither of them has a QC. At this time, the replicas may be unaware of the timeout because of network latency or a malicious leader. In addition, this study presented a view-change solution to guarantee that these two unsafe blocks are forked and a new block is extended at least from the locked position, such that all the locked blocks are eventually committed after the GST.

MPH was implemented in GO and deployed with up to 58 nodes in a LAN and WAN environment (with different bandwidths). Thereafter, it was compared to the state-of-the-art BFT protocols such as HotStuff and Streamlet, which have a pattern similar to the protocol proposed in this paper. The results showed that MPH convincingly outperformed HotStuff in terms of throughput by $60\%$ and outperformed Streamlet by more than $60\%$, regardless of the network condition. Moreover, the performance under a faulty leader or network instability was also comparable to and sometimes exceeded that of the state-of-the-art protocols.

The remainder of this paper is organized as follows. Section 2 describes the system model and preliminaries. Section 3 presents an analysis of the DiemBFT, a variant of HotStuff, and explores the factors contributing to performance metrics. Sections 4 and 5 provide the design and implementation of the protocol, respectively. Section 6 presents the safety and liveness proofs. Further, the evaluation and related works are presented in Sections 6 and 7, respectively. Finally, the conclusions and future works are presented in Section 8.

\section{Preliminaries}
This study assumed a system comprising a fixed set of $n=3f+1$ replicas, each replica indexed by number $i$, where $i\in \{1,...,n\}$. Replicas corrupted up to $f$ of the total by the adversary are referred to as Byzantine faulty and may arbitrarily deviate from the protocol, while the rest of the replicas are correct. Specifically, adaptive and static adversaries were considered for the upper and lower bounds, respectively. However, the difference was that \emph{static} type of adversary must decide the replicas to be corrupted at the beginning of every run, whereas \emph{adaptive} can decide the same during operation.
\subsection{Commnunication and Network}
As a distributed system, Network communication links are reliable and authenticated. They are implemented on a point-to-point basis; that is, all messages sent among correct replicas are eventually delivered if and only if the sender sent that message to the receiver. However, the links can be controlled by the adversary; the adversary controls the delivery time. This study referred to the network assumption proposed by Dwork et al. \cite{10.1145/42282.42283} to model the links, where there is a known bound $\Delta$ and an unknown Global Stabilization Time (GST). For instance, the messages transmitted between two correct replicas arrive within time $\Delta$ after GST, which is referred to as a \emph{partially synchronous}. In contrast, there is no guarantee that the message will arrive with $\Delta$, that is, the network is \emph{asynchronous}.
\subsection{Cryptographic Assumptions}
A standard digital signature was assumed, and public-key infrastructure (PKI) \cite{8809541,10.1007/3-540-45682-1_30,10.1007/3-540-45539-6_15} was provided by a trusted dealer. All replicas were equipped with a single public key, and each of the $n$ replicas was assigned a distinct private key. A replica $i$ can use its private key to sign a partial signature on message $m$, which is denoted by ${\rho}_i \gets tsign_i(m)$. The $(k,n)$-threshold signature scheme, a set of partial signatures $P=\{\rho_i|{\rho}_i \gets tsign_i(m),i\in I,|I|=k\}$ of any $k$ replicas can be used to generate a digital signature $\sigma \gets tcombine(m,P)$ on message $m$, where $I$ is the set of replica indexes that contributed partial signatures. Further, any replica can verify the signature using the public key through the function $tverify(m,\sigma)$. Moreover, a replica is aware that quorum replicas have given valid partial signature on message $m$ if the $tverify(m,\sigma)$ return \emph{true}. Although perfect cryptographic schemes do not exist in practice focusing on the distributed aspect of the problem in a cryptographic is sufficient. Consequently, given a valid signature $\sigma$ on $m$, no adversary or adversaries has an overwhelming probability of generating a signature $\sigma$ for $m$. This study adopted the threshold of $k=2f+1$, where the valid signature $\sigma$ on message $m$ implied that quorum replicas in such a system had received the message $m$.

In addition, a cryptographic hash function $h(\cdot)$ \cite{10.1007/978-3-540-25937-4_24}, which mapped an arbitrary length input to output of fixed size was assumed. The probability of any replica generating a couple of distinct inputs $m$ and $m^{'}$ subject to $h(m)=h(m^{'})$ is negligible; and the output of the hash function is referred to as the digest of the input, which can be used as a unique identifier for input.
\subsection{BFT SMR}
In a distributed environment, the consistency of all replicas is obtained through State Machine Replica \cite{10.1145/98163.98167,10.5555/296806.296824}, wherein each replica begins with the same state and receives inputs in the same order to reach the same state. The BFT protocol aids in each correct replica safely committing transactions from the client in a total order manner. It is expected that the protocol never compromises the safety, and the following property must be guaranteed in all runs:
\begin{itemize}
    \item Safety: All correct replicas commit the same transaction at the same log position.
    \item Liveness: Each transaction from client is eventually committed by all correct replicas. 
\end{itemize}
Regarding validity, the committed transactions satisfying certain application-dependent predicate can be implemented by the addition of validity \cite{10.1007/3-540-44647-8_31} checks on transactions before replicas propose or vote. For simplicity, a valid ``transaction domain'' was considered as a binary case, which is the value $v\in V, V=\{0,1\}$. 
\begin{itemize}
    \item Validity: if all correct replicas propose the same value $v\in V$, then no correct replicas commit a value other than $v$.
\end{itemize}
The SMR problem solved by HotStuff \cite{10.1145/3293611.3331591} provides the new pattern to be followed, which involves pipelining the different voting stages of the same batch of transactions into the one-round voting of different batches. Consequently, this study also followed the path to solving the BFT Agreement problem, with further details presented in section 3.
\subsection{Notations}
First, the notations used throughout the whole protocol are elucidated.

\textbf{View Number.} The BFT protocol is implemented via the preceding instance of single-shot Byzantine agreement individually; at most, one block can be committed within one instance. Further, each replica indexes the instance by the view number, which is initially to 0 and then increases monotonically.

\textbf{Block structure.} The block is denoted by a tuple $b=<id, v, p, txs, qc,\rho>$, where $id=hash(v,p,txs,qc)$ is the digest of the current block's information, which is also used as block's identification. $v$ is the view number of block; each block must refer to the block id in the previous view as the precursor that indicates which block the current block is based on (precursor field is used in our protocol). Further, $txs$ is a batch of pending transactions, $\rho$ is the partial signature from the proposer on block's $id$, and $qc$ is also defined blew.

\textbf{Quorum Certificate} A Quorum Certificate(QC) is formatted as a tuple $qc=<block,v,\sigma>$.It is a data type for proof where the block $B$ is validly signed on block id, generated via the combination of the partial signatures from a quorum $n-f=2f+1$ replicas. Consequently, it is implies that a block is received by a quorum replica and verified. Given a data structure of $x$, $x.y$ was used to refer to the element $y$ of the original structure $x$ throughout this study. In addition, the blocks and QCs can be ordered based on their view number; the highest block or QC indicates that the view number is the largest.

\textbf{Timeout Certificate} A timeout mechanism is required such that the system obtains liveness under partial synchrony. A timeout certificate (TC) was generated through the combination of a quorum of $n-f=2f+1$ timeout message, the partial threshold signature of timeout view $v$, and the highest QC $qc_h$ of replica's local. In contrast to QC, TC must contain the respective $qc_h$ in the timeout message from $n-f$ replicas, denoted by $tc=<v,qc_{high},\sigma>$, where $qc_{high}$ is the QC with the highest view number of its sender.

\subsection{Complexity Measurement}
We measure the complexity of protocol using the number of authenticators instead of message size because the message size is not related to metrics of the cost for one transaction. In addition, the bit of message cannot be bounded well to capture the cost. However, every message must be signed by the sender, and the receiver must verify the signature and combine quorum partial signature shares; these cryptographic operations are computationally intensive. Thus, the authenticator complexity is a suitable way to capture the overall costs of the protocol. 
\section{Description of Pipeline Paradigm}
This section introduces a variant of HotStuff referred to as DiemBFT. There are two components of the protocol, with the idea being inspired by PBFT \cite{10.5555/296806.296824}, linear \emph{normal commit path} progression, and quadratic cost \emph{view-change} owing to the malicious leader or asynchronous periods. During the normal process, a designated leader $L$ proposes a block $b$ that extends the block associated with the highest $QC$ of its own. Thereafter, other replicas attempt to update their locked view and the highest QC to check if any block can be committed when receiving block $b$. Then, block $b$ is voted for by sending a partial signature to the next leader $L_{v+1}$. Subsequently, the next leader $L_{v+1}$ forms a QC for view $v$ when it collects $n-f$ votes of the previous view $v$ and enter view $v+1$. Consequently, a new block is proposed by a leader for that view. The normal case repeatedly drives the system to commit transactions linearly; each block extends a block of the previous view, with a QC of that extended block being sufficient to prove safety. However, in case of any timeout owing to delay of the transmission or a malicious leader remaining silent, the timeout message being broadcast in view $v$ must carry a QC, which is the highest QC of the sender. This results in the new leader possessing the ability to prove that no action more recent than the block of that QC is committed. The leader proposes a block with TC of view $v$, where the TC contains a threshold signature on view $v$. Further, the proposed block in view $v+1$ extends the block with the highest QC in the timeout messages. When any replica receives a block with a TC, it attempts to advance the view and checks if any block satisfies the 3-chain commit rule, which implies that the first of three adjacent blocks that have QC and all uncommitted blocks before the first can be committed.

\textbf{Vote Rules}. Before a replica vote for the proposal, whether at least one of two rules is satisfied is examined as follows:
\begin{itemize}
    \item $b.v=b.qc.v+1$ 
    \item $b.v=b.tc.v+1$ and $b.qc.v \geq qc_{locked}.v$
\end{itemize}
Block $b$ contains the QC of the previous view or TC, which was formed in the previous or bad views. The first rule implies that voting for a block with a monotonically increasing view is safe and directly extends the block of $b.qc.block$, owing to no possibility of two QCs forming in one view. If a block $b'$ is certified except the case when a block that has QC and $b'$ s.t. $b.v=b'.v+1$, $f+1$ correct replica votes for $b'$ and updates its $qc_h$ to QC of block $b'$. The second rule implies that a valid block must extend at least a locked block, which is $qc_{high}.block$. Once a bad view occurs, at least one QC $qc_{high}$ of block $b$ must be included in any future timeout message. Moreover, even if one of the certain logging replicas proposes a block with a QC, all correct replicas' vote rule predicate can be true, although a certain few are ahead of the leader; thus, the block $b$ will never be forked.

\subsection{Complexity}
From a theoretical perspective, the 3-chain commit rule brings higher efficiency under synchrony and correct leader because multicast communication is disassembled into two linear broadcast communication through the introduction of the threshold signature scheme. Considering the complexity in \emph{view-change} phase, the three-phase consensus mechanism eliminates the requirement if replica broadcasting its own messages set of ``Prepare'' to prove its state in the view-change process from PBFT. In contrast, the timeout message is sent to a new leader by each replica. The communication complexity of all cases in HotStuff remains linear, such as $O(n)$. A block requires seven rounds of message exchange to be committed, containing two views and a QC broadcast phase; bringing network assumption into the rule implies that the latency of one commit is $7\Delta$.
\subsection{Throughput and Latency Analysis}
Considering the study on dissecting the chained-BFT performance \cite{9546436}, the life-cycle of the manner in which a block can be certified was analyzed to obtain the correlation between the permanence metrics and the vital environment factors. In a distributed setting, data in any form that is transmitted over the network must be bit stream such that the bandwidth is the upper limit of task processing in distributed cooperative systems. In pipelined methods, a consensus instance comprises proposing and voting steps, where the time consumption of an instance can be used to calculate the throughput and latency. A consensus instance begins with building a proposal, then transmitting it to every other replica, and ends when the leader of the next view collects quorum votes on the previous proposal such that a QC is formatted to advance the view to the next. 

Here, the focus is on the synchronous network, while Byzantine faults are ignored. The average service time that a block is certified can be divided into several parts:
\begin{equation}
  t_s=3t_{CPU}+2t_{NIC}+t_L+t_{Q}
\end{equation}

where $t_{CPU}$ captures the delay of signature operations (can be a constant number), $t_{NIC}=2mb^{-1}$ is the delay of data frame conversion rate at sender and receiver, where $m$ is the total size of a block, $t_L$ is the Round-Trip Time \cite{10.1145/3299869.3319893} that captures the network transmission delay by the normal distribution, and $t_{Q}$ is the expected delay of collecting a quorum of votes from replicas, which is modeled by $(\frac{2N}{3}-1)$ order statistics of $N-1$. As is known, each new certified block commits an ancestor certified block, and $t_s$ is used to approximately compute TPS, which is denoted as $TPS\approx|m|_{tx}t_s^{-1}$. Further, the BPS can be obtained as
\begin{equation}
  BPS\approx\frac{1}{(3t_{CPU}+2t_Q+t_L)*m^{-1}+4*b^{-1}}
\end{equation}
However, this is far from reaching the upper limit of bandwidth despite the first term of $(3t_{CPU}+2t_Q+t_L)*m^{-1}$ being close to 0. Thus, regardless of $b$ value, the overall $BPS$ provided by any distributed cooperative system is much smaller than the bandwidth, particularly when $3t_{CPU}+2t_Q+t_L$ contributes in excess. In addition, the average latency can be modeled as
\begin{equation}
  Latency\approx3t_{s}+6t_L
\end{equation}

\subsection{Limitations.} 
The core contribution of the Pipeline method is that unifying the different phase in PBFT's \cite{10.5555/296806.296824} consensus instance into a proposing-voting pattern. Recall the model given above, it provides a quantitative methodology to show which part of time consumption is contributing to the total block service time within one instance of Pipeline method. The term $t_{CPU}$ and $2t_{NIC}+t_L$ occurs in sequence, resulting in the same period of time $(\Delta=3t_{CPU}+2t_{NIC}+t_L)$, with either only CPU or network contributing to the actual block service time. First, it is embodied in the consensus that only one block is to be proposed in the two rounds of messages. Further, the period of time $2t_{NIC}+t_L$ is only used for sending or receiving messages and not for building proposals or preparing votes simultaneously. In particular, the network delay term significantly exceeds the CPU term. It is similar to one single pipe running such that other non-leaders are not fully utilized except for the period of process vote. Second, the safety of consensus requires that a view number can only be assigned to one block; only the leader was eligible to propose. Moreover, the advancement of a view must be after QC is formatted. However, the design of the single-pipe protocol itself becomes a performance bottleneck that is difficult to breakthrough.

\begin{figure}[ht]
  \centering
  \includegraphics[width=1\hsize]{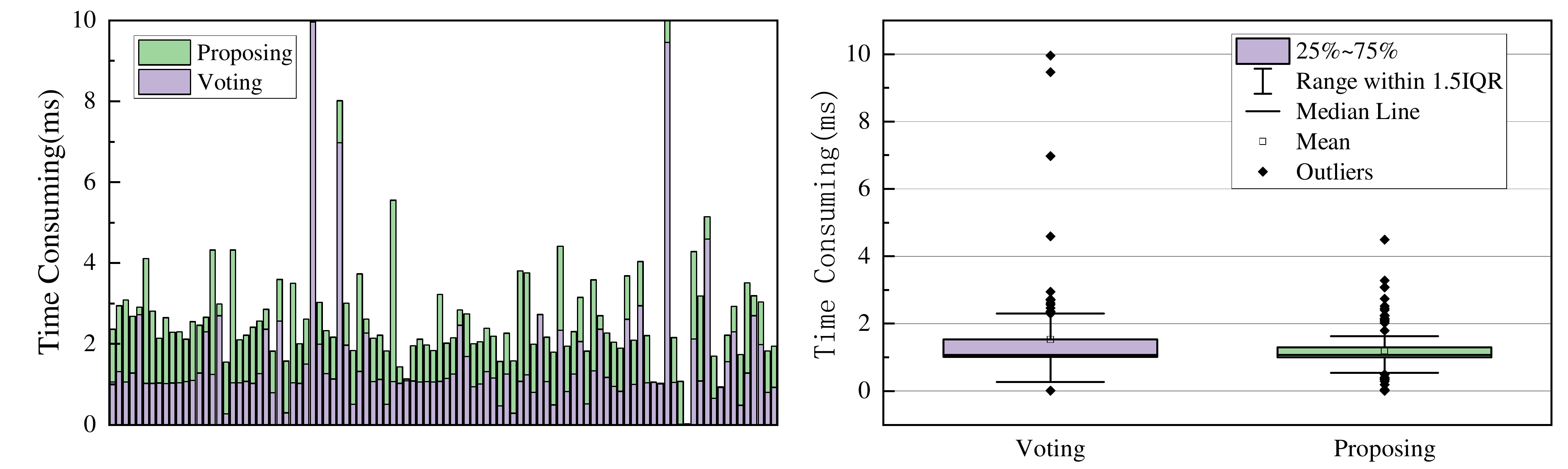}  \caption{time consumption of proposing and voting phase within one view}
  \label{fig:fig1}
\end{figure}

As shown in Figure \ref{fig:fig1}, the time required for both voting and proposing round during the running of implementation of the standard HotStuff was measured; the consumption of two rounds is approximately equal except for abnormal data, and each requires up to half of the time-consumption of each view. Further, the parallelism of the single pipeline is low as replicas are required to wait for another round of voting after proposing, thereby resulting in a lower block generation rate. Thus, the essence of the bottleneck is that the proposing and voting round is sequential in a single pipeline, thereby generating insufficient blocks. The next section shows that it is possible to better utilize each round by shrinking two rounds of the view into one round to exploit each round. The validator votes in each round, whereas the leader proposes in each round.

\section{M-Hotstuff Design}
To strengthen the performance of this pipeline method optimization in partial synchronous BFT protocol such as Hotstuff or its variants, we propose a multi-pipeline scheme that can make the replica propose and vote in every message round. We call the new protocol Multi-pipeline Hotstuff (MPH), which has linear communication cost for \emph{normal phase} and quadratic cost for \emph{view-change phase}. Although theoretically cost is the same cost as Hotstuff, the performance is almost doubled. In this paper, we only take two pipelines into consideration and leave exploration to future work, the full picture of protocol is presented in algorithm \ref{alg:alg1}. 

\begin{figure}[ht]
  \centering
  \includegraphics[width=\linewidth]{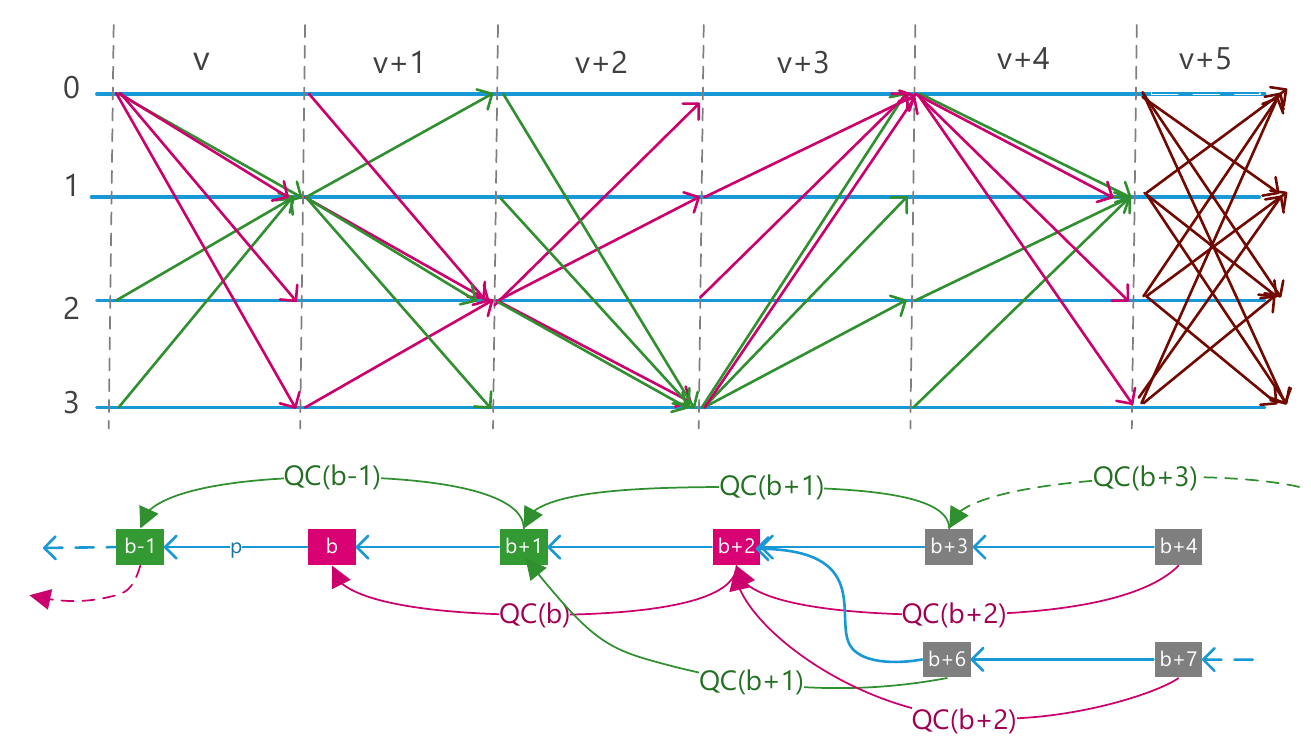}
  \caption{Normal Phase Overview of Multi-pipeline Hotstuff}
  \label{fig:fig2}
\end{figure}

\textbf{Protocol Overview} Different from the original works within one view, other replicas do nothing, only run the timer when the leader proposes. On the contrary, the idea behind our multi-pipeline is that the correct leader optimistically makes a proposal that extends the latest block, which is considered to be safe to vote. At the same time, the none-leader (validator) makes vote for block in that view. Note that, at most one replica has more than one role in view. In other words, voting and proposing happen at the same time but for different blocks. Such as view $v+2$, the leader of $v+2$ proposes a block $b+2$, the validators vote for block in view $v+1$. As we can see from Figure \ref{fig:fig2}, each view contains only one round of message exchange for voting and proposing; more blocks will be proposed in the same number of rounds compared to the original single-pipeline implementation. In MPH, each block contains two ``links'' to previous block; one is block's id of the parent and indicates the global order of block. And the other is the QC of the previous block, which indicates that the previous block was voted by quorum replicas. The block can be committed once it is the first of three or two adjacent blocks linked by the QC ``link'' depending on the 3-chain or 2-chain rule from the Pipeline paradigm. 

\subsection{Complementary of model}
\textbf{Timeout Message.} the timeout message is used to form a complete TC, which is denoted by a tuple $tm=<vf,vs,\rho>$. Here $vf$ is the current view that time outs and is used to form a TC that used to advance the first view after view-change. Further, $vs$ is the next view of $vs$; and $\rho$ is the partial signature shared on $vs$ and $vs$. \\
\textbf{Block structure.} The block structure is defined in multi-pipeline pattern as $b=<id,v,p,pv,txs,qc,tcs,tc,\rho>$, where $tc$ and $tcs$ are current formed by replicas, $tc$ is used by those lagging replicas and $tcs$ is a set of TC, which will be used to verify the $tc$ is valid. Further, $pv$ is the view of block with ID $p$ and can be used to check safety rule when time outs.

\subsection{One Round Trip View}
MPH also implements BFT protocol through consensus instances indexed by view numbers. Each view has a designated leader, which is a replica deterministically defined by the view number (e.g., $id=v\bmod n $, where $n$ is the total number of replicas). The difference lies in the fact that only one round message trip occurs within a view, the leader makes a proposal and broadcasts it to other replicas, and then the view is updated through the QC in the proposal. As soon as the replicas receive the proposal, the validator votes for the corresponding block, which implies that voting and proposing to occur parallelly if a replica is leader. In contrast to HotStuff, the proposing and voting processes of the same block occurs sequentially within the same view, and the replicas update the view after the block has a valid QC in the current view. In MPH, owing to only one round trip occurring in each view, only one of the above events can be achieved. This ensures the continuous advancement of the view each round, and consequently, the view number of the block's QC is computed by $qc.v=b.v+1$.
\subsection{Block State}
Owing to the design of one round trip view to more precisely elaborate the whole protocol, two types of block states during the consensus process, the \textbf{certified} and \textbf{verified} states, were defined. A block is considered certified if there exists a QC for the block, that is, $qc.block=b$. Further, a block is verified if the proposer's partial signature was authenticated and the QC in the block was well-formatted by a quorum. The former state implies that a quorum of replicas have seen the block and voted for it. Thus, following the 3-chain commit rule, the block can be committed after another two certified blocks. A correct replica will perform the proposing and voting step based the current block state. The new certified block will commit a preceding certified block, and the new verified block will be extended to build a new proposal for next view.

\subsection{3-chain Predicate in M-Hotstuff}
While MPH performs in terms of the 3-chain rule, the protocol tracks certain vital variables of block and QC to satisfy safety and liveness. The first of two adjacent certified blocks chained by QC is referred to as the \textbf{locked} block, which indicates that no higher block could have reached a commit state. Further, the QC corresponding to the block is referred to as locked QC. Recall the \textbf{Vote Rule} referred to in Section 3, which stated that a proposal is accepted by correct replicas if the branch of the new proposal extends from the currently locked block and the view number is monotonically increasing. In this study, two types of locked variables were defined: $qc_{curlock}$ used to keep track of the current locked position to which pipeline the currently processing block belongs, and $qc_{laslock}$ used to keep track of the previous locked position where another pipeline exists. Because there are multiple locked positions in the multi-pipeline setting, a proposal is expected to extend at least a locked block, which is the block certified by $qc_{curlock}$. In addition, another two variables record the highest and the second-highest QC, denoted as $qc_{high}$ and $qc_{sechigh}$, respectively. These are used to form a new chained block for their respective pipelines. 
\begin{figure}[ht]
  \centering
  \includegraphics[width=0.75\hsize]{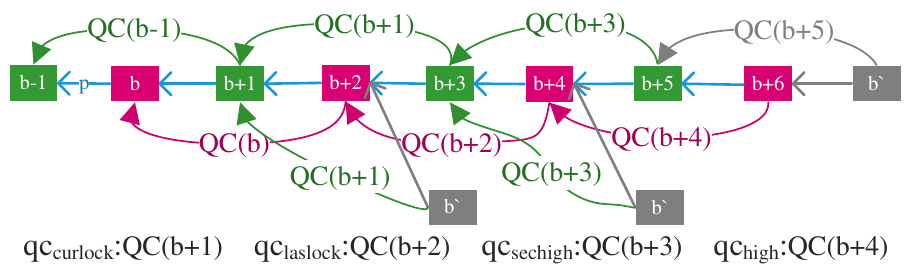}
  \caption{3-chain predicate of Multi-pipeline HotStuff}
  \label{fig:figp}
\end{figure}
\subsection{Normal-Case Operation}
In MPH, a transaction pool was designed to buffer transactions from the client to the provider proposer for fetching payloads in batches. However, as the size of the pool affects system latency, it was set as a configured parameter in subsequent evaluation. In a stable view, the Leader $L$, upon entering view $v$, collects votes for the proposal in the previous view $v-2$ to generate certification (QC, suppose $qc$) through the combining of quorum partial signature shares. Thereafter, a block $b$ extends the latest verified block that is known to the leader. According to the Block Structure defined in Section 2, the QC is included inside the block, and a batch of transactions is pulled from the pool as payload ($txs$). However, other metadata must be formatted correctly. Moreover, the leader no longer extends the certified block by the highest QC. Instead, it extends the latest verified block, which is essential to ensure that each round of view generates a new block.

As a validator, when receiving the first view $v$ block from $L$, any replica attempt to verify the authenticity of the leader's proposal $b$ to check if the proposer is the leader of the block's view and if the signature on the block is valid. Subsequently, the block's state becomes verified, and if the QC in block $b$ is valid, then the block associated with QC becomes certified, that is, $qc.block$. Thereafter, validators advance its view number, and the second-highest QC is updated to the highest QC, followed by the highest QC to the latest QC in block $b$. Moreover, the $laslock$ and $curlock$ are updated one step forward. Thus, batch transactions in the first block of three adjacent certified blocks are chained by QC, which are provided to the execution module to commit. Optimistically, a block that is considered verified will eventually be certified in the next view unless the next leader is faulty. Based on this, two parallel sub-procedure were designed for simultaneous proposing and voting. If the validator is the next leader, it sends a signal to the proposing procedure to start making a proposal for that view, provided there is a valid QC of the previous view generated. Meanwhile, the validator follows the 3-chain rules to check if this block can be voted. Consequently, according to the 3-chain predicate defined above, the safety rule for voting in the normal case is obtained as
\begin{itemize}
\item Safety: $b.v=b.qc.v+1$ and $b.pv=verifiedB.v$, the block $b$ directly extends from the latest verified block with latest QC.
\end{itemize}
If the proposal is satisfied with the safety rule, then a threshold signature on $b$ is sent to the leader of view $b.v+2$ as a vote. Finally, the normal case repeatedly drives the system to process transactions.

\subsection{View-Change Case}
When the timer of a certain view $v$ expires, owing to the lack of a mechanism to detect whether the timeout is caused by asynchrony, the message is delayed, or the leader is faulty. Consequently, the view-change mechanism is triggered to advance to the next view $v+1$ and replace another leader for ensuring liveness. In the case of this study, two blocks were considered uncertified. Consequently, the view-change period of MPH comprised two consecutive views, each of which required a corresponding TC be advanced. As the timeout occurs, all replicas stop voting for that view and multicast a timeout message containing a threshold signature share for $v-1, v$ and its second-highest QC $qc_{sehigh}$ as well as highest QC $qc_{high}$. When any replicas receive a quorum of such timeout messages, first, the timeout messages are checked to keep having the latest QC. Thereafter, two TCs of view $v$ and $v+1$ are formed through the combination of quorum partial signature shares, denoted as $tc1$ and $tc2$, respectively. Finally, the liveness rule for voting can be expressed as:
\begin{itemize}
\item Liveness1: $b.v=b.tc.v+1$ and $b.qc.v=qc_{curlock}.v+i*2, b.pv= qc_{laslock}.v-1+i*2,i\in\{0,1\}$.
\item Liveness2: $b'.v=b'.tc.v+1$ and $b'.qc.v=qc_{laslock}.v+i*2, i\in\{0,1\}$ and $b'.pv=verifiedB.v$.
\end{itemize} 

In the first view, after timeout, the leader of that view $v+1$ first enters through TC $tc1$, and then proposes a block that directly extends the block of $qc_{high}.block$. Recall that in the necessary metadata of the structure of the block defined in Section 2, the two TCs $tc1 and tc2$ and QC $qc_{sechigh}$ must be included inside the block. Once the validators receive the block from the leader, it updates its own view based on TC and checks the cryptographic semantics of the block. Further, it computes whether the block is a valid branch and changes the verified branch to the new block. The Liveness1 rule states that a valid proposal in the first timeout view must contain a QC that is at least as new as $qc_{curlock}$, and the block must directly extend from the successor block of $b.qc.block$. If Liveness1 is satisfied, then the same parallel sub-procedure is repeated; the next leader first updates its view through second TC $tc2$, and the vote of $b$ is sent to the leader of view $b.v+2$. When the leader reaches the new view $b.v+2$, it makes a proposal that directly extends from the latest verified block, where the proposal also contains two TCs and the latest QC $qc_{high}$. In contrast to the normal case, no votes are collected in this view. In the second timeout view, the validators also update the view number by TC $tc2$ and check the authenticity of the proposer of the block using the Liveness2 predicate to determine whether to accept the block. Here, the QC $b'.qc$ cannot be higher than the QC $qc_{curlock}$, and $b'$ must be a successor of latest verified block. Consequently, all replicas vote for $b'$, the next leader also receives votes of view $b.v$ while also proposing for the next view. 
As now a view-change process after the timeout has finished, it is guaranteed that the system will attempt to keep working in the event of timeout due to asynchrony or a faulty leader problem. Thus, it first needs to synchronize the view such that at least a quorum replicas start with the same view to prevent inconsistency problems caused by out-of-sync blocks. Regardless of normal or view-change cases, the voting predicate is true provided either one of the safety and liveness rules holds. Thus, any replica can use it to determine whether to vote for the proposal.

\subsection{Complexity and Latency}
Obviously, when the network is synchronous, and the leader is honest, a complete threshold signature of size $O(1)$ within QC needs at least quorum partial signature shares. Every share of size $(1)$ is given by a unique replica's cryptographic tool as a vote and sent to the leader. Owing to a constant number of committing a transaction, it can be concluded that the overall complexity of reaching a consensus decision in a normal case is $O(n)$. Thus, each transaction adopts the 3-chain to be committed, and one more round for all replicas receives the proof. This results in a latency of seven rounds. In the view-change case, when asynchrony occurs, all replicas need broadcast timeout messages with QC for view synchronization, and each message has size $O(2)$, and the subsequent view-change message of the leader's proposal contains three authenticators ($O(3)$), corresponding to one QC and two TCs. Therefore, the responding vote also has a single share of size $O(1)$. Consequently, the view-change case yields $O(n^2)$ and $O(n)$ complexities of view synchronization and view-change, respectively. For latency, in a locked transaction that suffered asynchrony with a latency of at least nine rounds, if the network returns to synchronous, there is at least one branch with two new blocks. The branch costs two more extra rounds. However, the round of latency goes to an infinite round if the GST never occurs.

\begin{algorithm}
\caption{multi-pipeline HotStuff for replica $i$}%算法名字
\label{alg:alg1}
\SetKwProg{InitialKw}{Initialize}{:}{}
\InitialKw(//initialize all relevant data and start protocol){}{
    $timer.Reset()$\;
    $currView\gets1$\;
    Send newView $M_{newView}$ to local
}
\SetKwProg{Fn}{Procedure}{:}{}

$\rhd$ \textcolor{blue}{Normal\ phase}

\Fn{block Process loop}{
    \While(:\ wait event $M$){$True$}{
        \lIf{$M$ is a proposal message}{$ProcessProposal(M)$}
    }
}
\Fn{proposing loop}{
    \While(:\ wait event $M$){$true$}{
        \lIf{$M$ is a newView message}{$Propose(M)$}
    }
}
\Fn{vote process loop}{
    \While(:\ wait event $M$){$true$}{
        \lIf{$M$ is a timeout message}{$ProcessTM(M)$}
        \lIf{$M$ is a vote message}{$ProcessVote(M)$}
    }
}
$\rhd$ \textcolor{blue}{Finally}

\Fn{timer loop}{
    \While(){$true$}{
        \If(//Check if the timer expires during any phase){$timer.stop()$}{
            $M.timeout\gets <vf:currView+1,vs:vf+1>$\;
            Broadcast timeout $M_{timeout}$ to all replicas
        }
    }
}
\end{algorithm}

\section{implementation}
This study implemented all protocols discussed: DiemBFT, a productive variant version of HotStuff and MPH in Golang. The TCP\footnote{\href{https://pkg.go.dev/net}{https://pkg.go.dev/net}} was used to build reliable point-to-point channels to implement the SMR-BFT abstractions correctly. Further, secp256k1\footnote{\href{https://pkg.go.dev/crypto/elliptic}{https://pkg.go.dev/crypto/elliptic}} was used for elliptic curve based signatures. Further, the implementation of Multi-pipeline\footnote{\href{https://github.com/tncheng/multipipelinehs}{https://github.com/tncheng/multipipelinehs}}, HotStuff and the corresponding single pattern protocol were open-sourced.

The main framework of the protocol is provided in Algorithm \ref{alg:alg1}, which is described as message-driven. All messages were generated to drive consensus in a view-by-view loop, and each type of message was given to the corresponding parallel procedure. Further, a replica performed phase in terms of the received message in succession based on its role, where the replica can have more than one role (the leader implemented using the round-robin-based way). The protocol was activated by a new-view message in the initialization step, and then all following runs were based on inner or external messages. In addition, a \emph{timer} mechanism always detected the network fails or message delays to ensure liveness.
\textbf{Ordering:} The set of committed blocks must be equipped with a total order and certified by three adjacent QC. Moreover, it must be a total ordered set. For instance, the $b.p$ gives the causal order, indicating that the block $b$ is directly extend from the block with id $p$. Further, the block's view number also can be used to describe the relation; for example, blocks with large view numbers are generated after blocks with smaller view numbers. For agreement, every replica should retain the same ordered pairs of the block. Considering the design of multi-pipeline, the unique ordered set may be obstructed by two obstacles, one is the ``location'' where the branch is built in any case, and the other is the monotonicity of consecutive advancement of the view when voting and proposing in parallel.
\textbf{Proposing:} Similar to the optimization presented in Section 4, the number of transactions processed by the protocol per unit time can be improved through chaining and adopting an optimistic path for proposing in each round of messages. There must be two uncertified blocks at any view. A normal commit path is that the new proposal carries the QC of the grandparent block and extends the parent block, implying that the leader proposes and votes in time, and nothing goes bad. However, once the timer detects the failure of the network or leader, the protocol must drop the latest two uncertified blocks and make a new branch from the latest certified block. Further, the corresponding QC embedded in the block should certify the block where the previous of the current extends from. In short, one normal and two types of proposing after timeout entails that the multi-pipeline guarantee the safety of BFT-SMR. The predicate for three types of proposing way is presented in Algorithm \ref{alg:alg3}, which guides the protocol to extend and certify block correctly.

\begin{algorithm}
\caption{utilities for protocol}
\label{alg:alg2}
\SetKwProg{Fn}{Function}{:}{\KwRet}
\Fn{QC(V)}{
    $qc.\sigma\gets tcombine({vote.\rho|vote\in V})$\;
    $qc.v\gets vote.v+1:vote\in V$\;
    $qc.bid\gets vote.block.id:vote\in V$\;
    \KwRet $qc$\;
}
\Fn{TC(TM)}{
    $tcf.\sigma,tcs.\sigma\gets tcombine(timeout.\rho|timeout\in TM)$\;
    $tcf.v,tcs.v\gets timeout.vf,timeout.vs+1:timeout\in TM$\;
    \KwRet $\{tcf,tcs\}$
}
\Fn{ProcessQC(qc)}{
    \lIf{$currView<qc.v$}{ \KwRet}
        $currView\gets qc.v+1$\;
        $secHighQC,highQC\gets highQC,qc$\;
        $timer.Reset()$\;
}
\Fn{ProcessTC(tc,tcset)}{
    \lIf{$(tc=\perp)\lor(tcset=\perp)$}{\KwRet}
    \ElseIf{$(currView<tc.v)\land(tc.v=min_v\{t.v|t\in tcset\})$}{
        $currView\gets tc.v+1$\;
        $secHighTC,highTC\gets argmin_v\{t.v|t\in tcset\},argmax_v\{t.v|t\in tcset\}$
    }
    \ElseIf{$(currView=tc.v)\land(tc.v=max_v\{t.v|t\in tcset\})$}{
        $currView\gets tc.v+1$\;
        $secHighTC,highTC\gets \perp,\perp$
    }
    $timer.Reset()$\;
}
\Fn{Verify(b)}{
    \lIf{not $tverfy(b,b.\sigma)$}{\KwRet $false$}
    \lIf{not $tverfy(b.qc,b.qc.\sigma)$}{\KwRet $false$}
    \KwRet $true$
}
\Fn{VoteRule(b)}{
    \lIf{$b.v\ne currView+1$}{\KwRet $false$}
    \lIf{$b.tc\ne\perp$}{ \KwRet safety is true}
    \ElseIf{$b.tc.v=min_v\{t.v|t\in b.TcSet\}$}{
        \KwRet liveness1 is true
    }
    \ElseIf{$b.tc.v=max_v\{t.v|t\in b.TcSet\}$}{
        \KwRet liveness2 is true
    }
    \KwRet $false$
}
\end{algorithm}
\textbf{View increasing:} At a high level, the synchrony of the network is used to model the transmission latency instead of the ordering of the arrival of messages. Hence, even a message sent first may arrive after messages sent later. The proposing and voting design in multi-pipeline are logically parallel, as opposed to sequentially proposing and voting in HotStuff. Consequently, the order in which votes of the previous view and the new proposal are received by a replica is not consistent with the order in which they should be. Some replicas may receive votes of the previous view and generate QC before the current block. Subsequently, the leader, if any, creates a new branch that conflicts with the unreceived block. The view number should be monotonically increasing even if the replica observes a QC that does not belong to the current view. A predicate of consecutive advancing view is presented in Algorithm \ref{alg:alg3}, which prevents block conflicts caused by inconsecutive QC.

In addition, the goal of the protocol is to reach an agreement on the order of transactions instead of the content itself. In the implementation, a subsystem of mempool is built, which removes the transaction dissemination from the critical path of consensus. Here, all the received transactions from clients are disseminated in batches instead of one by one. Therefore, the transaction body need not be included in consensus blocks. In addition, the consensus only agrees on the digest of these transactions, implying that all the replicas' mempool is shared. This significant optimization has been proven by Narwhal \cite{10.1145/3492321.3519594}, which can further improve performance compared to other studies that have not equipped the sharing strategy. However, in a slight deviation from Narwhal, each digest of the transaction is included in the consensus messages rather than a simple digest of a batch of transactions. This facilitates the determination of the order for multiple batches in every consensus instance.

\begin{algorithm}
\caption{main implementation of Multi-pipeline HotStuff}
\label{alg:alg3}
\SetKwProg{Fn}{Procedure}{:}{}
% \SetKwProg{FnInterrupt}{Function}{:}{\KwRet}
\Fn{ProcessProposal}{
    \lIf{not\   $Verify(M.block)$}{\KwRet}
    $ProcessQC(M.block.qc)$\;
    $ProcessTC(M.block.tc)$\;
    $B^*\gets M.block;B'\gets B^*.qc.block;B''\gets B'.qc.block;B'''\gets B''.qc.block$ \;
    \If{$(B^*.qc.bid=B'.id)\land (B'.qc.bid=B''.id)\land(B'qc.v>curlockQC.v)$}{
        $laslockQC,curlockQC\gets curlockQC, B'.qc$
    }
    \If{$(M.block.tc\ne\perp)\land(M.block.TcSet\ne\perp)\land(M.block.tc.v=min_v\{t.v|t\in M.block.TcSet\})\land(i=LeaderOf(M.block.v+1))$}{
            $M_{newView}.Vtype\gets TimeoutS$\;
            Send newView $M_{newView}$ to local
    }
    \If{$(B^*.qc.bid=B'.id)\land (B'.qc.bid=B''.id)\land(B''.qc.bid=B'''.id)$}{
        Commit $B'''$, reply
    }
    \lIf{not\ $VoteRule(M.block)$}{\KwRet}
    $verifiedB\gets max_{v}\{verifiedB,M.block\}$\;
    Send vote $M_{vote}$ to $LeaderOf(M.block.v+2)$ 
}
\Fn{Propose(M)}{
    \lIf{$i\ne LeaderOf(currView)$}{\KwRet}
    \If{$M.Vtype=Normal$}{
        $b \gets$ \textcolor{blue}{extends $verifiedB$ with $highQC$}
    }
    \If{$M.Vtype=TimeoutF$}{
        $b \gets$ \textcolor{blue}{extends $highQC.block$ with $secHighQC$ and $\{secHighTC,highTC\}$}
    }
    \If{$M.Vtype=TimeoutS$}{
        $b\gets$ \textcolor{blue}{extends $verifiedB$ with $highQC$ and $\{secHighTC,highTC\}$}
    }
    $M.block\gets b$\;
    Broadcast Proposal $M_{proposal}$ to all replicas
}
\Fn{ProcessVote(M)}{
    $V\gets V\cup M.vote$\;
    \If{$(|V|=n-f)$}{
        $qc\gets QC(V)$\;
        Wait until $currView=qc.v+1$ to consecutive advance view\;
        $ProcessQC(qc)$\;
        $M.Vtype\gets Normal;M.tc=\perp$\;
        Send newView $M_{newView}$ to local\;
    }
}
\Fn{ProcessTM(M)}{
    $TM\gets TM\cup M.timeout$\;
    \If{$|TM|=n-f$}{
        $M.TcSet\gets TC(TM)$\;
        $ProcessTC(argmin_v\{t.v|t\in M.TcSet\})$\;
        $M.Vtype\gets TimeoutF$\;
        Send newView $M_{newView}$ to local
    }
}
\end{algorithm}

\section{Correctness Proof}
Recall that as per the definition of BFT protocol in Section 2.3, it must satisfy two properties, safety, and liveness. If a block is committed by a correct replica, then all other correct replicas should eventually commit the same block in the same view, and subsequently, the committed block must form a linear chain linked by their respective QC. In this Section, the proof of the safety and liveness of multi-pipeline HotStuff is presented.
\subsection{Safety}
We start with some definition that we will use:

\begin{itemize}
    \item $b_i\gets*b_j$ indicates that the block $b_j$ extends the block $b_i$.
    \item $b_i\gets qc_{i+1}\gets b_{i+2}$ indicates that the block $b_i$ is certified by the QC of $qc_{i+1}$, which is contained in the block $b_i+2$.
    \item $prepareQCView(b_i):=b_{i-4}.v$, such that $b_{i-4}\gets qc_{i-3}\gets b_{i-2}\gets qc_{i-1}\gets b_i$.
    \item $lockedView(r,b_i)$ can return the locked view of replica $r$ after voting for block $b_i$.
\end{itemize}
Again, we denote an element $y$ of data structure $x$ as $x.y$.\\
\textbf{Lemma 1.} Given any valid two QCs, say $qc_1$ and $qc_2$, there must be $qc_1.v\ne qc_2.v$. \\
\emph{Proof.} We prove this by contradiction, suppose there exist $qc_1.v=qc_2.v$. The section 2 tells that a valid threshold signature in QC can be formed only with $n-f=2f+1$ partial signatures for it, and each correct replica only votes once in the same view. Consider that, the set $R_1$ is the replicas that have voted for block $qc_1.block$, subject to $|R_1|=2f+1$. Likewise, $N_2$ is for $qc_2.block$, subject to $|R_2|=2f+1$. Since the model of system is $n=2f+1$, we have $|R_1|+|R_2|-n=1$, by the quorum intersection, we can conclude that there must be a correct replica who voted twice in the same view. It's contradict with the assumption, $qc_1.v=qc_2.v$ is proved. \\
\textbf{Corollary 1.} By the Lemma 1, If $qc_1,qc_2$ are two valid QCs, such that $qc_1.v\ne qc_2.v$, by the quorum intersection, $R_{qc_1}\cap R_{qc_2}=r$, then there must exist a correct replica $r$ such that $r\in qc_1$ and $r\in qc_2$. \\
\textbf{Lemma 2.} Consider two certified blocks: $b\gets qc$ and $b'\gets qc'$. under BFT model, if $b.v=b'.v$ then $b=b'$. \\
\emph{Proof.} By Lemma 1, we have two valid QCs must be $qc.v\ne qc'.v$, unless $qc=qc'$. Two certified blocks has the same view, by the rule of $VoteRule$ and $ProcessQC$, there must be $qc=qc'$, then the Lemma 2 is true. \\
\textbf{Lemma 3.} Assuming that a chain consists of three adjacent certified blocks, which is linked by QC, and start at view $v_0$ and end at view $v_4$. For every certified block $b\gets qc$ such that $b.v>v_0$, then we have that $prepareQCView(b)>v_0$. \\
\emph{Proof.} Let $b_0\gets qc_{b_0}\gets b_2\gets qc_{b_2}\gets b_4\gets qc_{b_4}$ be that chain starting as view $b_0.v$ and ending view $b_4.v$. By Corollary 1, the intersection of QC $qc_2$ and $qc$ has at least one correct replica, denote as $r$. Since $b.v>v_4$ and $VoteRule$, $r$ must votes on $b_4$ first, then update it's lock on $b_0$, that is, $lockedView(r,b_4)=v_0$. Because the locked view never decreases, when $r$ votes for the block after $b_4$, such as $b$, the voting rule of $r$ implies that $prepareQCView(b)>v_0$. \\
\textbf{Lemma 4.} Assuming that a chain with three contiguous view starting with a block $b_0$ at view $v_0$. For every certified block $b\gets qc$ such that $b.v\geq v_0$, then we have that $b_0\gets*b$. \\
\emph{Proof.} Again, let $b_0\gets qc_{b_0}\gets b_2\gets qc_{b_2}\gets b_4\gets qc_{b_4}$ be the chain starting with $b_0$ and other block's view is contiguous, such that: $v_0+4=b_0.v+4=b_2.v+2=b_4.v$. There are two cases:
\begin{itemize}
    \item If $v_0\leq b.v\leq v_0+4$, then $b.v$ can be one of the values: $v_0,v_0+2,v_0+4$. By the Lemma 2, $b$ is one of blocks: $b_0,b_2,b_4$. Hence, $b_0\gets*b$.
    \item If $b.v>v_0+4$. By the Lemma 3, we have $prepareQCView(b)>v_0$, this means there exist a block before $b$, which is certified by $b.qc$, say $b_i$, such that $b_i.v\geq v_0$. Since $v_0\leq b_i.v<b.v$, we could apply same induction hypothesis on $b_i$ de deduce that $b_0\gets*b_i$. Finally, $b_0\gets b_i\gets b$ concludes the proof of $b_0\gets*b$.
\end{itemize} 
\textbf{Lemma 5.} Consider two conflict blocks: $b_0$,$b'_0$, such that $b_0.p=b'_0.p$ and $b_0\ne b'_0$, only one of them can be committed by a correct replica. \\
\emph{Proof.} We proof the by contradiction, suppose two blocks can be committed, then each block is extended by two certified block. Let $b_0\gets qc_{b_0}\gets b_2\gets qc_{b_2}\gets b_4\gets qc_{b_4}$ and $b'_0\gets qc'_{b'_0}\gets b'_2\gets qc'_{b'_2}\gets b'_4\gets qc'_{b'_4}$ are the commit chain of block $b_0$ and $b'_0$ respectively. By Lemma 2, $b_0\ne b'_0$ unless $b_0.v=b'_0.v$, we assume that $b'_0.v>b_0.v$. By Lemma 4, we must have $b_0\gets *b'_0$, this is contradict with $b_0.p=b'_0.p$ and $b_0\ne b'_0$. Therefore, Lemma 5 is true.

\subsection{Liveness}
\textbf{Lemma 6.} When a replica in view $v-1$ receives a proposal for view $v$ from another correct replica, it advances into view $v$. \\
\emph{Proof.} If a correct replica make a proposal for view $v$, it must have seen votes for $v-2$ or timeout messages for $v-1$, and well formated a QC or TC of view $v-1$. When a correct receives such a proposal with QC or TC, it will advance the view and enter view $v$. \\
\textbf{Lemma 7.} After GST, the message delays between correct replicas are finite, then all correct replicas keep their view monotonically increasing. \\
\emph{Proof.} suppose that all correct replicas start at least view $v$, and let $R$ is the set of correct replicas in view $v$. There are two cases of set $R$: 
\begin{itemize}
    \item If all $2f+1=|R|$ correct replicas time out in view $v$, then all replicas in $R$ will receive $2f+1$ timeout messages to form a TC and enter view $v+1$.
    \item Otherwise, at least one correct replica $r'$, not having sent timeout message for view $v$. $r'$ must have observed a QC of view $v-1$ and keep updated its $qc_{high}$ accordingly. If $r'$ times out in any view$>v$, then the updated $qc_{high}$, which is never decreased, must be contained in timeout message, it will trigger $R$ to enter a view higher than $v$. Otherwise, $r'$ must see a QC in all view$>v$. by Lemma 6, the proposal which contains that QC from correct replica will eventually be delivered to $R$, triggering it to enter a higher view.
\end{itemize}
\textbf{Lemma 8.} If correct replicas has just entered view $v$, no QC has yet been generated and none of them has timed out. when correct replica $r$ receives a proposal from correct leader in view $v$, $r$ will vote for the propposal. \\
\emph{Proof.} the predicate in $VoteRule$ checks that:
\begin{itemize}
    \item View numbers are monotonically increasing. By assumption, none of them has timed out means that no TC and QC could have been generated for view $v$ and $v-1$. Therefore, when predicate $VoteRule$ execute on a proposal from correct leader is the first largest voting view.
    \item If TC in proposal is empty, the proposal must directly extends the latest verified block in view $v-1$ and in which QC must directly extends must have view $v-2$, all view is consecutive. 
    \item If TC in proposal is non empty, it is based on $2f+1$ timeout messages. A correct leader must track the latest QC to have view at least as large as the $qc_{high}$ of each timeout messages during collecting it. The predicate on QC in proposal must satisfy with $qc_{*lock}.v-b.qc.v\in\{0,2\}$ and $qc_{high}.ve>qc_{*lock}.v$. Apparently, $r$ send a vote for the proposal from correct leader.
\end{itemize}
\textbf{Lemma 9.} After GST, at most $2\Delta$ from the first replica entering view $v$, all correct replicas receive the proposal from the correct leader. \\
\emph{Proof.} When the first correct replica enters view $v$, if it is leader, all the correct replicas will receive proposal within $\Delta$; if not, it must have a TC for view $v-1$. Next, it will forward TC to the leader of view $v$ within $\Delta$. Upon the leader enters view $v$, it immediately makes a proposal and multicasts it to other correct replicas within $\Delta$. \\
\textbf{Lemma 10.} After GST, for every correct replicas in view $v$ eventually certify a block $b$ and commit a block $b_c$, such that $b.v>v,b_c\gets qc_{b_c}\gets *b\gets b.qc$. \\
\emph{Proof.} Recall that our system model defined in section 2, at most $f$ out of $n$ are faults, such that $n>3f+1$. Each leader of view designated by round-robin, let $n=4,f=1$, which is mean that there are three adjacent correct primaries, say $v',v'+1,v'+2,v'>v$. By Lemma 7, all correct replicas keep advancing view. By Lemma 9, at most $2\Delta$ from any correct replica enters view $v'$, the proposal of $v'$ will be delivered to all correct replicas. By lemma 8, correct replicas vote for the proposal from correct leader. Repeatedly deducing the argument by lemma 7,8,9, the block of view $v',v'+1,v'+2$, say $b_{v'},b_{v'+1},b_{v'+2}$, will append to the chain until next faulty leader comes up. Where at least one($b_{v'}$) of three new blocks will be certified by QC, which is contained in third block of view $b_{v'+2}$. By the rule of $ProcessQC$ and $VoteRule$, when a correct replica successfully perform the vote step on block of view $v'+2$, the first block $b_c$ among new three adjacent certified blocks linked by QC, will match with commit rule: $b_c\gets qc_{b_c}\stackrel{certified\  blocks}{\longleftarrow}b_{v'}\gets qc_{b_{v'}}\gets b_{v'+1}\gets qc_{b_{v'+1}}\gets b_{v'+2}$. Therefore, the block $b_{v'}$ can be certified and $b_c$ can be committed. Each message in view must be delivered within $\Delta$, all correct replica receives proposal of view $v'+2$ after at most $4\Delta=2\Delta+\Delta+\Delta$ time of the first correct replica entering view $v'$.

\section{Evaluation}
This section evaluates the throughput and latency of the proposed implementations through experiments on WAN (Alibaba Cloud) and LAN. The leader aim was to demonstrate the better performance of the multi-pipeline mechanism compared with that of HotStuff and other consensus protocol that have similar principle, such as Streamlet. The multi-pipeline mechanism fills batches of transactions per round of message through separated proposing and voting, which utilizes the gap time in a single pipeline to transport more transactions and efficiently utilize bandwidth. However, the commit rule in the multi-pipeline mechanism also follows the 3-chain rule, which can maintain the same level of commit latency as HotStuff. 

A testbed on LAN containing 12 local machines, each supported by Intel Xeon Gold 6230 processors (20 physical core), 256 GB of memory, and a Debian 11 server, was deployed. For a large-scale test, virtualization technology was employed to provide up to five virtual instances per physical machine to run replicas. The maximum TCP bandwidth measured by iperf was approximately 1 GB per second. In addition, the WAN used an ecs.g6.xlarge instance across 5 different regions: Zhangjiakou, Chengdu, Qingdao, Shenzheng, Hangzhou. They provided a maximum bandwidth of 12.5 MB, 4 virtual CPUs on a 2.5 GHz, Intel Xeon(Cascade Lake) Platinum 8269, and 8 GB of memory, while running on Debian 11.

Throughput and end-to-end latency were utilized as performance metrics. Throughput is an overall metric that measures the number of transactions that can be committed per second, denoted as TPS. Further, end-to-end latency measures the average time required for a transaction to be submitted to the system until it is committed, including network transmission time, waiting time in transaction buffer, and validation time. As mentioned, the mempool implementation undertakes the transaction body transmission ability for a fairer measurement of the performance. All the compared protocols had the exact shared mempool implementation.

\subsection{Best Performance}
First, the throughput and latency were measured on a small scale, (4, 10 replicas) replicas and 800 block size (digests). Here, each payload size of the transaction was set to 1024 Bytes, and the mempool's dissemination batch size was set to 512 KB. The network maintains continuous synchronicity and triggers no view-change.
\begin{figure}[ht]
  \centering
  \subfigure[]{\includegraphics[width=0.49\hsize]{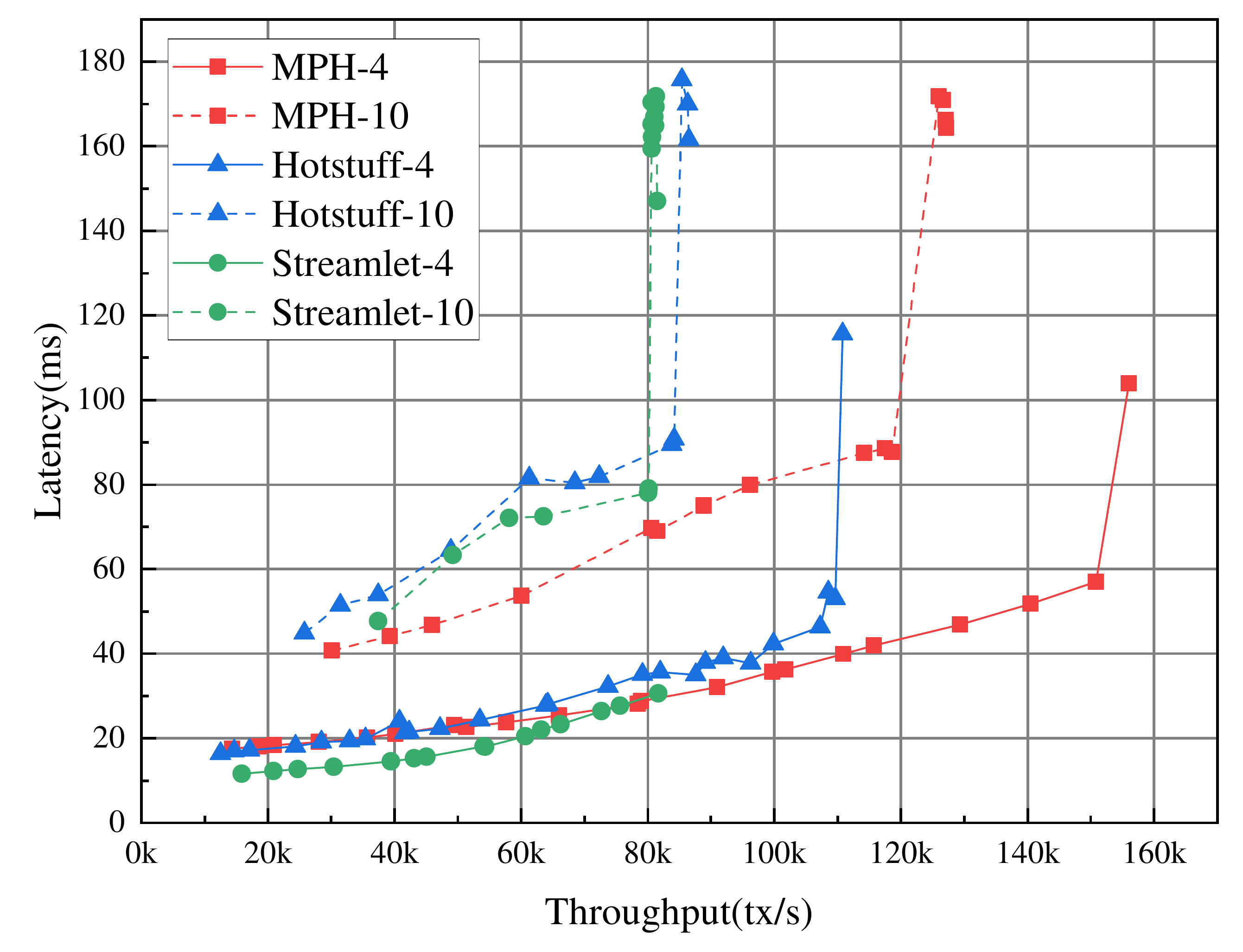}\label{fig:fig3:a}}
  \subfigure[]{\includegraphics[width=0.49\hsize]{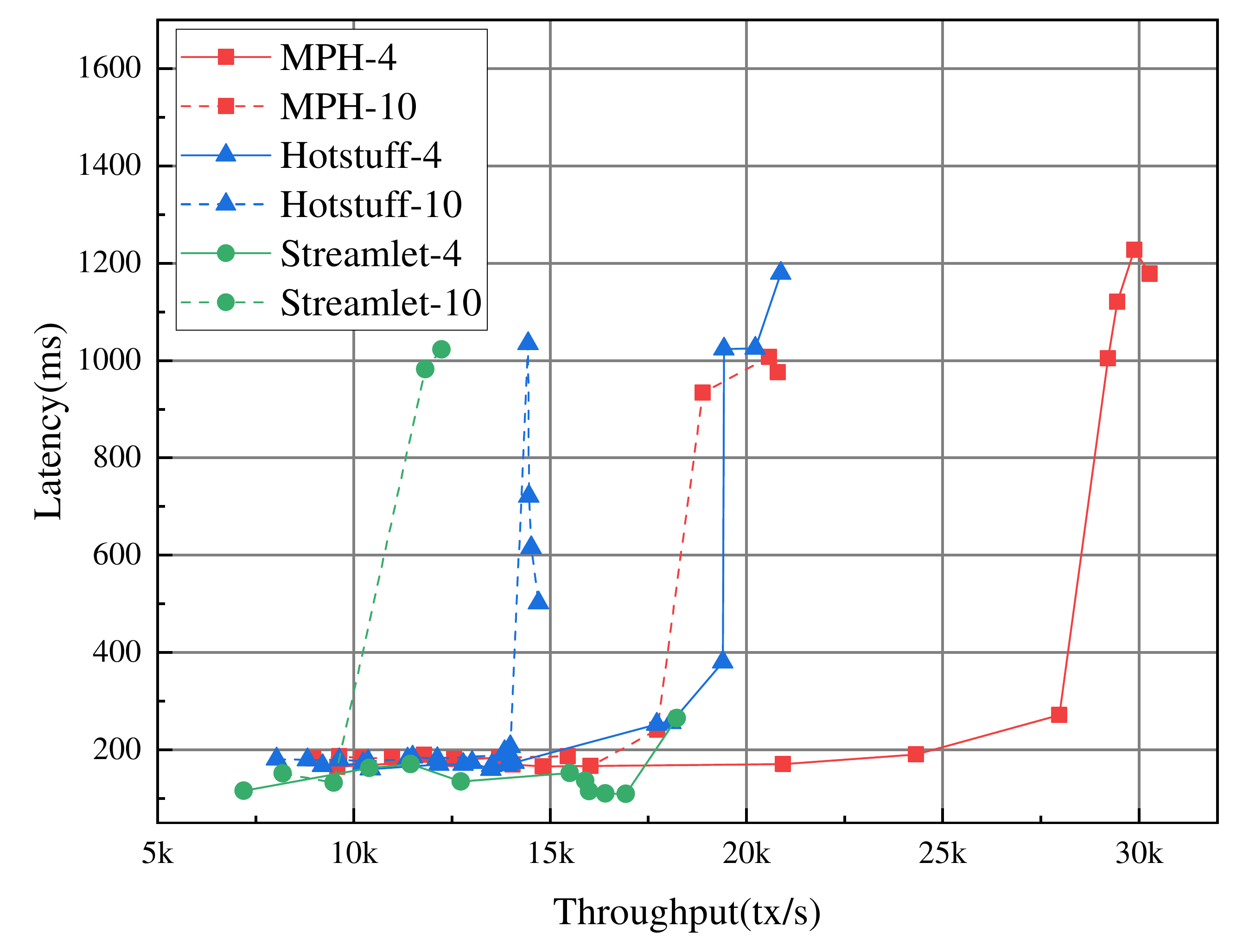}\label{fig:fig3:b}}
  
  \caption{Throughput vs. latency with network size 4,10 and batch size 800, on LAN and WAN}
  \label{fig:fig3}
\end{figure}

Figure \ref{fig:fig3} depicts the trend of latency with an increase in TPS under the setting of two network conditions and two network sizes. Different transaction arrival rates were set to obtain different TPS when all replicas conducted correct behavior, and the network was synchronous. As can be observed from Figs. \ref{fig:fig3:a} and \ref{fig:fig3:b}, the setting 4 and 10 replicas were denoted as ``-4'' and ``-10.'' Further, the LAN (\ref{fig:fig3:a}) and WAN (\ref{fig:fig3:b}) network condition results showed that regardless of network conditions, the proposed protocol can break through performance bottlenecks and significantly improve TPS with approximately the same latency. As observed from both trends, the latency of the proposed protocol was very close to that of HotStuff, and even the TPS increased up to approximately 160k on LAN. Owing to limited network bandwidth and long-distance transmission, the maximum TPS was only 30K on WAN, while the latency was greatly increased.

Under the same network environment, each committed block in multi-pipeline underwent the same certified process as HotStuff; thus, they maintained the same latency level. When the network size was 4, the latency of Streamlet was minimal and increased slowly with TPS. As the size scaled up to 10, the latency became larger than multi-pipeline. The maximum throughput of multi-pipeline exceeded that of HotStuff and Streamlet by approximately $50\%$ and $100\%$, respectively. In addition, benefiting from the concurrent replicas and higher utilization of bandwidth, it can process more transactions in the same number of steps. However, the throughput of Streamlet is worse than both in this setting, owing to its quadratic messages for each operation, including proposing and voting. The quadratic communication complexity consumes more bandwidth and burdens the system performance as the system scales up. 

\subsection{Scalability}
The scalability of three protocols with system scales of 4 to 58 was evaluated, where all the experiments were performed in the same environment with no extra network delays. A 1024B-size payload, 512KB-size dissemination batch, and a maximum 800-number transaction(digest) were used as the system configuration. Further, the transaction arrival rate was set to be maximum until the system saturated to record the maximum throughput and latency while varying the number of replicas. In addition, to capture the error bars of each setting, the standard deviation was calculated for ten runs with the same setting.
\begin{figure}[ht]
  \centering
  \subfigure[]{\includegraphics[width=1\hsize]{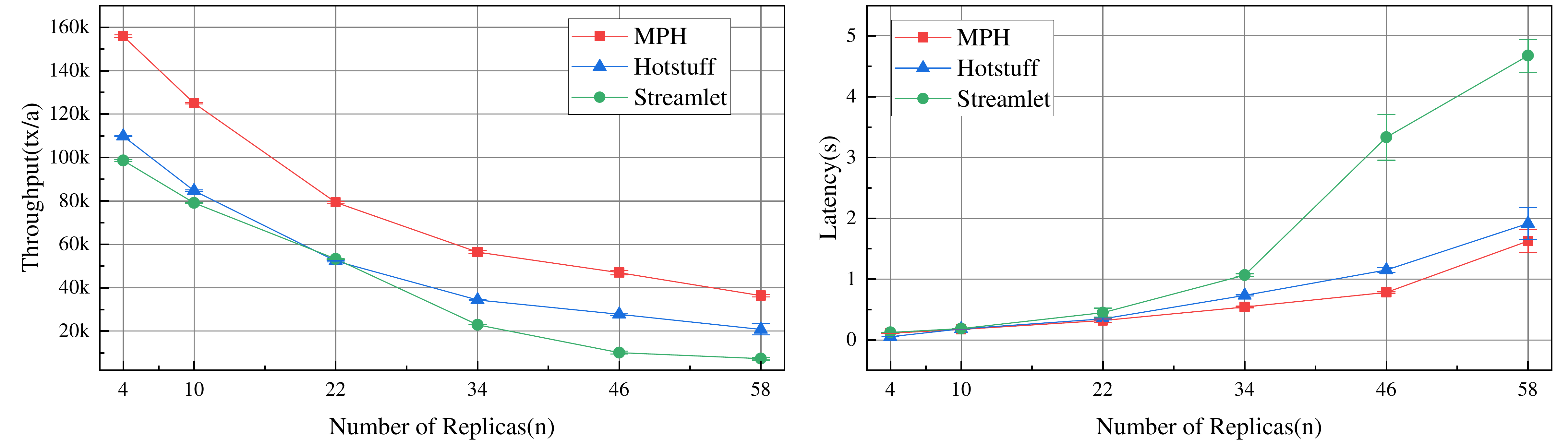}\label{fig:fig4:a}}
  \subfigure[]{\includegraphics[width=1\hsize]{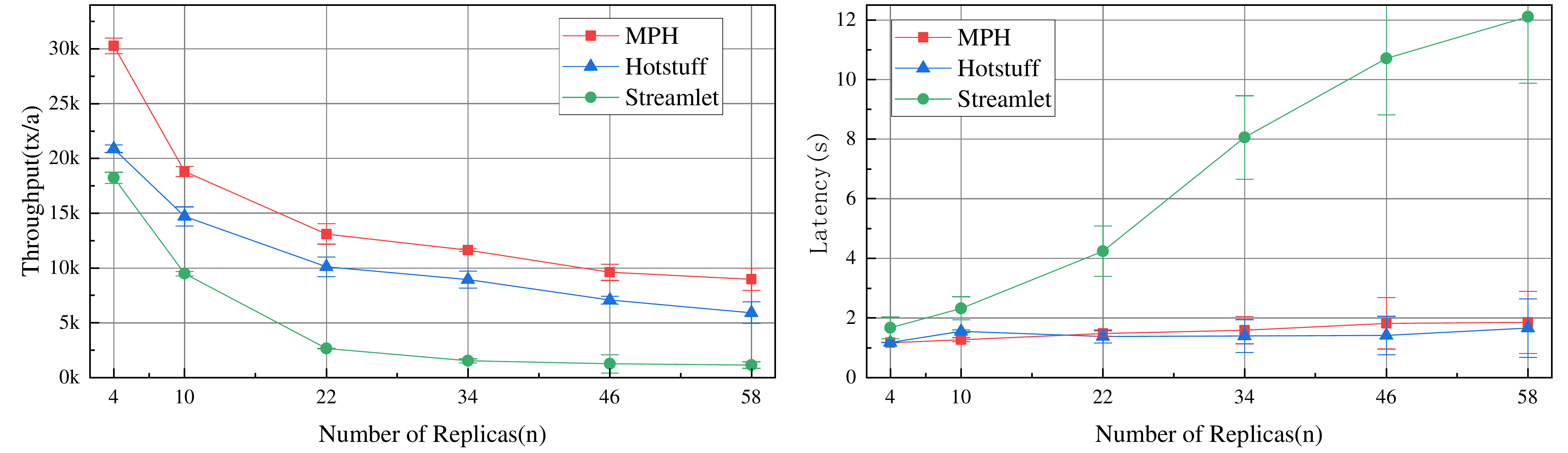}\label{fig:fig4:b}}
  \caption{Scalability with batch size 800 on LAN(a) and WAN(b)}
  \label{fig:fig4}
\end{figure}

Figure \ref{fig:fig4} shows the throughput and latency over the number of replicas scale-up. As can be observed, the latency of MPH is very close to that of HotStuff, and approximately $60\%$ better throughput than HotStuff at each system scale on LAN \ref{fig:fig4:a} and WAN \ref{fig:fig4:b}. The throughput of Streamlet drops sharply from the same level as HotStuff at the beginning owing to the quadratic communication cost. In general, when the system scales up to 58 in the LAN, the MPH provides nearly 40k TPS while maintaining an average latency of 2 s. Further, the linear communication cost allows HotStuff and MPH to maintain a slow latency growth, whereas, in Streamlet, it increases dramatically. Owing to the transmission delay in the WAN, the latency of all protocols was higher than the LAN environment at each system scale and with a slower increase than the LAN environment except Streamlet. However, the MPH still provided approximately 10k TPS with a maximum latency of 3.5 s, which was better than HotStuff.
\subsection{View-change}
Three protocols with 22 replicas were executed with certain being set as faulty (0,1,2 or 3 crashed) at the beginning of the experiment on LAN, denoted by ``-1,-2,-3.'' In addition, payload and dissemination sizes were set to 1024B and 512KB, with maximum 800 block size. Moreover, the timeout was set to 500 ms. Owing to the use of leader rotation in pipeline pattern to provide fairness, this configuration forces the protocol to perform frequent view-change. Therefore, the experiments exhibited the degree to which view-change slowed down the performance of consensus. 
\begin{figure}[ht]
  \centering
  \includegraphics[width=1\hsize]{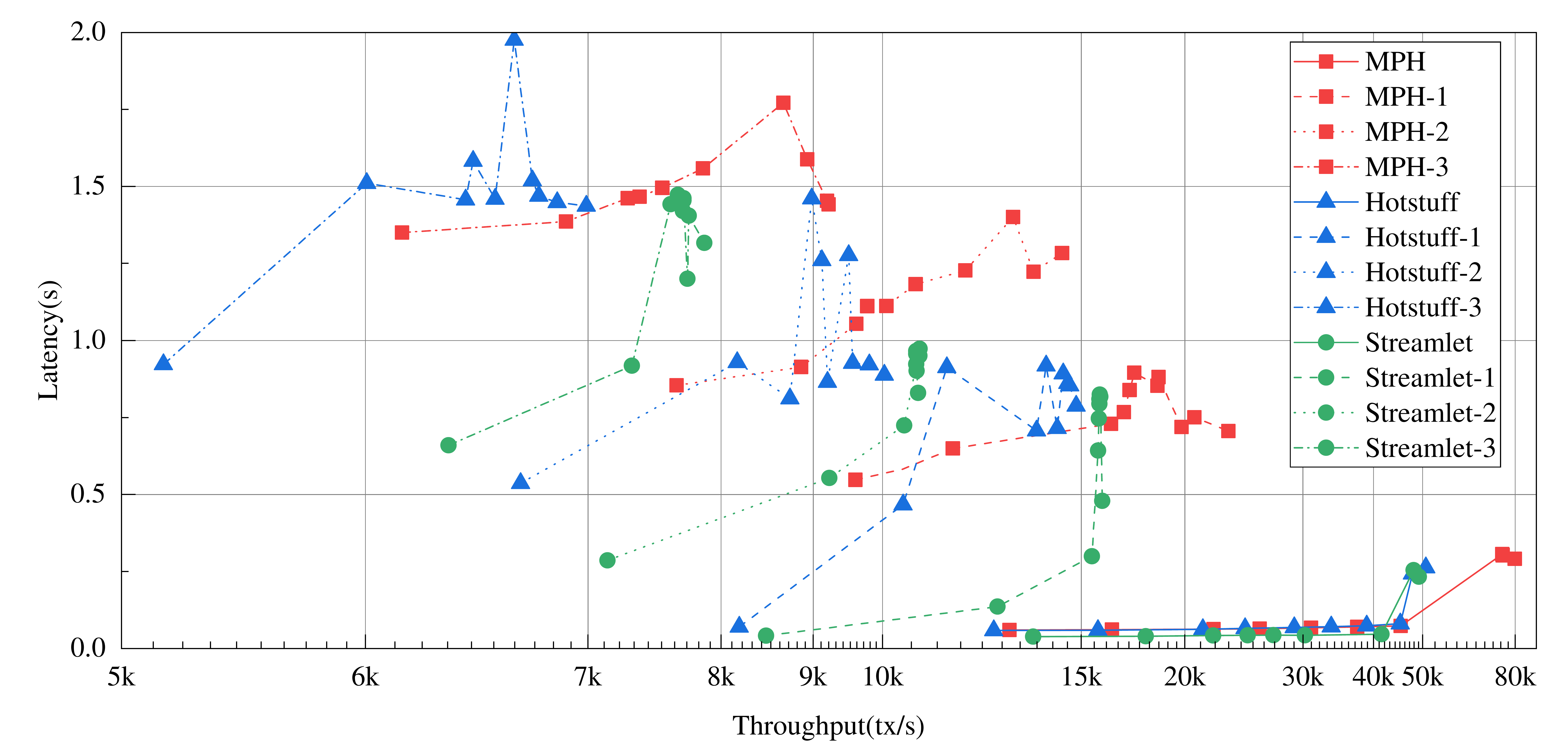}
  \caption{Throughput vs. latency with network size 22 and batch size 800}
  \label{fig:fig5}
\end{figure}
As evident from Figure \ref{fig:fig5}, MPH outperformed HotStuff and Streamlet at the beginning, and with increase in the number of faulty replicas, the frequent view-change rendered it closer to Hostuff. Because the feature of MPH inherently improves the performance than HotStuff, the throughput loss of MPH for the same number of faulty replicas was lower than other two. Starting from 30k improvement with no faulty replicas to 2k improvement with 3 faulty replicas, the MPH performance gains compared to HotStuff and Streamlet gradually tapered off. The disappearance of multi-pipeline optimization is primarily because of t, and, where more forked blocks than a single pipeline must be processed (at most 4 blocks will be forked in the proposed implementation, twice as many as HotStuff) and only three consecutive views without malicious replicas can continue to effectively drive consensus after the timeout (assuming view $v$ is the first view after the timeout, even a correct leader of view $v+2$ can collect a quorum votes of view $v$, the new QC will never be observed by other replicas once the leader of $v+1$ is faulty; new view-change will be triggered by view $v+1$). As for latency, the maximum latency of MPH increased from 200 ms to approximately 2 s as more replicas became faulty. When TPS reached its peak, MPH and HotStuff performance were at the same level; however, it was slightly higher than that of Streamlet. The reason why MPH performed better than HotStuff is that more TPS significantly reduced the waiting time in the buffer. It is worse in the case of Streamlet because Streamlet has no view synchronization phase to perform because it follows the longest branch extending rule regardless of the phase. Moreover, the timeout in Streamlet was fixed. 

\section{Related work}
State Machine Replication abstracts the operation on multiple replicas into a whole. This abstraction relies on consensus protocol at its core to order the requests from clients, such that they are committed in the same order. Reaching consensus in the presence of Byzantine failures was abstracted as Byzantine Generals Problem by Lamport et al. \cite{10.1145/3335772.3335936}, who also introduced the term ``Byzantine Faults,'' a semantics of distributed failures. Several previous studies on fault-tolerance did not involve Byzantine faults or only considered an ideal synchronous network (e.g., \cite{10.1145/3335772.3335934,10.1145/3335772.3335939}). To solve crash fault-tolerance problems using SMR, Lamport proposed Paxos \cite{lamport2001paxos,10.1145/3335772.3335939}, a leader-based protocol that drives consensus decisions in the pipeline based on the designated leader, with linear communication complexity and one round message exchange. However, Paxos is difficult to understand. Consequently, a simpler Raft \cite{184040} was devised by Diego and John, which decomposed the consensus problem into three independent components and provided a better foundation for system building. 

Tolerating Byzantine faults requires a much more complex protocol, which suffers from expensive message complexity or time-consuming authenticator steps. The first theoretical synchronization solution in a fault-tolerant system was provided by Pease et al. \cite{10.1145/322186.322188}, wherein an exponential number of messages was sent to reach an agreement. Thereafter, the first practical polynomial algorithm was proposed by Dolev, and Strong \cite{10.1145/800070.802215,10.1145/2455.214112}, with the tolerance of $f<n$. The communication complexity was $O(n^3)$ in any case, with or without authentication. To further improve efficiency, the randomized method was introduced \cite{4568104,doi:10.1137/S0097539790187084,10.1145/872035.872066,10.1007/11818175_27,10.1007/978-3-030-32101-7_20}. Kztz and Koo \cite{10.1007/11818175_27} proposed an expected constant-round and $f<n/2$ resilience protocol using key primitive referred to as moderated Verifiable Secret Sharing (VSS) and later improved by Abraham and et al. \cite{10.1007/978-3-030-32101-7_20}. The improved protocol has expected $O(1)$ round complexity and an expected $O(n^2)$ communication complexity. Recent studies of Abraham et al., Sync HotStuff \cite{9152792} have shown a simple and practical BFT solution that achieves consensus with a latency of $2\Delta$ in the normal case(where $\Delta$ is the upper bound of message delay) and quadratic communication. Another study \cite{cryptoeprint:2021:1138} allowed optimistic responsive leader rotation in addition to shrinking Sync HotStuff's latency upper and lower bounds.

The partial synchronous model was first proposed by Dwork, Lynch, and Stockmeyer(DLS) \cite{10.1145/42282.42283}, who also first proposed a protocol that preserved safety before GST and guaranteed liveness following GST costing $O(n^4)$ communication and $O(n)$ round per decision. However, expensive costs resulted in the protocols remaining theoretical until PBFT was devised by Castro and Liskov \cite{10.5555/296806.296824}, which is an efficient protocol. PBFT implements BFT SMR using reliable broadcast with $O(n^2)$ communication complexity and two round-trips per decision, in addition to $O(n^3)$ messages in view-change case. Several studies have been conducted to reduce the communication cost, with quadratic cost in the normal and view-change cases being referred to as Tendermint \cite{kwon2014tendermint} having two round commit latency. Kotla et al., introduced Zyzzyva \cite{10.1145/1323293.1294267}, which considered that an optimistic path has a linear cost, while the view-change path remained $O(n^3)$. Thereafter, a safety violation was found by Abraham et al., who also presented revised protocols \cite{abraham2017revisiting}. In the best case, the over-quadratic cost was the bottleneck that rendered the deployment of the system on a large scale challenging. SBFT \cite{8809541} was shown to address the challenge of scalability based on key ingredients: using threshold signatures to reduce communication to linear and an optimistic path to reduce latency. Consequently, the state-of-the-art protocol was HotStuff \cite{10.1145/3293611.3331591}, which bridges the world of BFT consensus and blockchain, and has $O(n)$ cost and optimistic responsiveness. These features, along with the threshold signature scheme and pipeline operation, have provided a new pattern for designing the modern protocol. From this, a variation of HotStuff, such as Fast-HotStuff \cite{jalalzai2020fast}, provided less latency through the reduction of 3-chain structures to 2-chain one. Further, the DiemBFT \cite{baudet2019state} from Meta further refined HotStuff as a core of the blockchain system. It retained linear costs and introduced an explicit liveness mechanism, which is its leader reputation mechanism that provides leader utilization under crash, thereby preventing unnecessary delays. Inspired by past studies, Chan et al. described an extremely simple and natural paradigm with quadratic cost referred to as Streamlet, \cite{10.1145/3419614.3423256}, which finalized a record when it was extended by another two certified records.

Moreover, FLP \cite{10.1145/3149.214121} theory stated that reaching deterministic consensus is impossible in the face of even a single failure. Fortunately, the impossibility was circumvented by harnessing randomness; several studies have focused on asynchronous settings, including the first practical asynchronous protocol given by Cachin et al. \cite{10.1007/3-540-44647-8_31}, HoneyBadgerBFT \cite{10.1145/2976749.2978399}, VABA \cite{10.1145/3293611.3331612}, Dumbo-BFT \cite{10.1145/3372297.3417262}, and DAG-Rider \cite{10.1145/3465084.3467905}. Most of these protocols have over-quadratic costs; however, they are always alive even in asynchrony.
\section{Conclusion and Future Work}
This study proposed MPH, the first protocol that improved performance without duplication problem caused by parallel proposing. It achieved consensus by through the combination of two pipelines in HotStuff into one instance to overcome the performance bottleneck when there are no faults. In combination with optimistically proposing, this allowed each leader to produce a block within at most $\Delta$ time after GST, further reducing the waiting time for block generation. Consequently, MPH achieved unprecedented throughput and outperformed state-of-the-art protocols. Based on the 3-chain predicate in safety checking, the end-to-end latency does not significantly increase. The evaluation conducted showed that MPH outperformed other protocols with a proposing-voting pattern in terms of throughput and latency, regardless of the network conditions.

An interesting future work is the independent processing of two HotStuff instances when time outs, which can further improve the throughput and latency performance of the system when there are exits faults. Further, more HotStuff instances may be combined with the multi-pipeline pattern for further optimization.

\bibliographystyle{unsrt}  
\bibliography{references}

\end{document}